\documentclass[epj]{svjour}
\usepackage{graphics}
\usepackage{amsmath,amssymb}
\usepackage{verbatim}
\def\rr{\mathbf{r}}
\def\kk{\mathbf{k}}
\def\be{\begin{equation}}
\def\ee{\end{equation}}
\def\bea{\begin{eqnarray}}
\def\eea{\end{eqnarray}}
\def\vn{\mathbf{0}}

\begin{document}
\title{Number of closed-channel molecules in the BEC-BCS crossover}
\author{F. Werner\thanks{\emph{Present address:} Department of Physics, University of Massachusetts, Amherst, MA 01003, USA}%
 \and L. Tarruell\thanks{\emph{Present address:} Institute for Quantum Electronics, ETH Z\"urich,
   8093 Z\"urich, Switzerland }%
 \and Y. Castin }
%
%
\institute{Laboratoire Kastler Brossel, Ecole Normale Sup\'erieure, UPMC and CNRS,
24 rue Lhomond, 75231 Paris Cedex 05, France}
%
%
\abstract{
Using a two-channel model, we show that the number of closed-channel molecules
in a two-component Fermi gas close to a Feshbach resonance is directly related to the derivative of the energy of the gas with
respect to the inverse scattering length. 
We extract this quantity from
the fixed-node Monte~Carlo equation of state and we compare
to the number of closed-channel molecules
measured in the Rice experiment with lithium 
[Partridge et al., Phys. Rev. Lett. 95, 020404 (2005)].
We also discuss the effect of a difference between the trapping potentials seen by a closed-channel molecule and by an open-channel pair of atoms
in terms of an effective position-dependent scattering length.
} 
\maketitle
\section{Introduction}
\label{sec:introduction}

It is now experimentally possible to prepare two-component atomic Fermi gases 
at low temperature with a fully controlable interaction strength, in a regime
where the interaction range is negligible and the atomic interactions are
characterized by a single parameter, the $s$-wave scattering length $a$
between two opposite ``spin" atoms.
The value of $a$ can be adjusted at will
thanks to a magnetically induced Feshbach resonance.
The weakly interacting limits $k_F a\to 0^+$ and $k_F a\to 0^-$,
where $k_F$ is the Fermi wavevector of the gas, correspond respectively
to the BEC limit (a Bose-Einstein condensate of dimers, observed in \cite{BEC_Inns,BEC_JILA,BEC_MIT,SalomonCrossover,Hulet}) 
and the BCS limit (a ``condensate" of Cooper pairs approximately
described by the BCS theory). Furthermore, the experiments can access
the so-called crossover regime between BEC and BCS, where the gas
is strongly interacting ($k_F|a| \gtrsim 1$), which includes
the celebrated unitary limit $k_F |a|=\infty$, where the gas acquires fully universal 
properties \cite{Thomas1,Jin,Salomon1,GrimmCrossover,Grimm_gap,ThomasVirielExp,GrimmModes,Ketterle_vortex,JinPotentialEnergy,thomas_entropie,HuletPolarized,Ketterle_unba}.

In this context, early theoretical studies \cite{Timmermans_phys_rep,Holland0,Holland} put forward a many-body
Hamiltonian which accurately models the  microscopic two-body physics
of the Feshbach resonance. The interaction potential is described by
two channels, an open channel and a closed channel. The atoms
exist in the form of fermionic particles in the open channel and
in the form of bosonic short range molecules in the closed channel.
Two atoms may be converted into a short range closed-channel molecule and {\sl vice versa}
due to the interchannel coupling. The corresponding Hamiltonian is
\begin{multline}
H_2 = \int d^3r \left\{\sum_{\sigma=\uparrow,\downarrow}
 \psi_\sigma^\dagger(\rr)\left[-\frac{\hbar^2}{2m}\Delta + U(\rr) \right] \psi_\sigma(\rr)
 \right.\\
\left.+
\psi_b^\dagger(\rr)\left[E_b(B)-\frac{\hbar^2}{4m}\Delta + U_b(\rr) \right] \psi_b(\rr)
\right\}\\
+
\Lambda \int d^3r_1 d^3r_2\, \chi(\rr_1-\rr_2) 
\left\{
\psi_b^\dagger[(\rr_1+\rr_2)/2]
\psi_\downarrow(\rr_1) \psi_\uparrow(\rr_2)
+ {\rm h. c.}
\right\}
\\
+ g_0 \int d^3r_1 d^3r_2 d^3r_3 d^3r_4 \,
\chi(\rr_1-\rr_2) \chi (\rr_3-\rr_4) \\ \delta\left(\frac{\rr_1+\rr_2}{2}-\frac{\rr_3+\rr_4}{2}\right)
\psi_\uparrow^\dagger(\rr_1) \psi_\downarrow^\dagger(\rr_2)
\psi_\downarrow(\rr_3) \psi_\uparrow(\rr_4)
\label{eq:definition_H2}
\end{multline} 
where the atomic fields $\psi_{\sigma}(\rr)$ obey the usual fermionic anticommutation
relations and the field $\psi_b(\rr)$ describing the closed-channel molecules
obeys the usual bosonic commutation relations.
$\Lambda$ gives the coupling between the closed and open channels,
with a short range cut-off function $\chi(\rr)$, of range $b$, and such that $\int d^3 r\, \chi(\rr)=1$. A typical value
for $b$ is in the nanometer range
\footnote{
$b$ can be estimated by the van der Waals length, that is the length that
one forms with $\hbar$, $m$ and the $C_6$ coefficient of the van der Waals
interaction \cite{Koehler_review}. 
}.
The coupling constant $g_0$ represents the background atomic interaction
in the open channel, which is conveniently modeled by a separable potential with
the same cut-off function $\chi$.
The energy $E_b$ is the energy of a closed-channel molecule, in the absence
of the coupling $\Lambda$, counted with respect to the dissociation limit
of the open channel. This energy is adjusted with a magnetic field $B$,
using the fact that the different magnetic moments in the open and closed
channels lead  to a differential Zeeman shift. The
resulting effective magnetic moment is
\be
\mu_b\equiv\frac{dE_b}{dB} .
\ee
$U(\rr)$ and $U_b(\rr)$ are the trapping potentials experienced by the atoms and
the closed-channel molecules respectively.

On the contrary, recent Monte~Carlo simulations of the many-body problem in the BEC-BCS
crossover simply use single channel  model Hamiltonians~\cite{DeanLee,bulgacQMC,zhenyaPRL,zhenyaNJP,Juillet,BulgacCrossover,zhenyas_crossover}. In most unbiased 
Quantum Monte~Carlo simulations, a lattice model is used:
\begin{multline}
H_1 = \sum_{\rr} b^3 \left\{\sum_{\sigma=\uparrow,\downarrow}
 \psi_\sigma^\dagger(\rr)\left[-\frac{\hbar^2}{2m}\Delta + U(\rr) \right] \psi_\sigma(\rr)
 \right.
 \\
 \left.
 + g_0 \psi_\uparrow^\dagger(\rr) \psi_\downarrow^\dagger(\rr) 
 \psi_\downarrow(\rr) \psi_\uparrow(\rr)\right\}.
\label{eq:H1}
\end{multline}
Here $b$ is the lattice spacing (which coincides with the ``range'' of the discrete delta interaction potential),
$\Delta$ is a discrete representation of the Laplacian, and $g_0$ is a coupling
constant adjusted to have the desired scattering length:
As was shown e.g.\ in \cite{zhenyaPRL,zhenyaNJP,YvanHoucheslowD,boite},
\be
\frac{1}{g} - \frac{1}{g_0} = 
\int_{[-\pi/b,\pi/b]^3} \frac{d^3 k}{(2\pi)^3} \frac{1}{2\epsilon_{\mathbf{k}}}
\label{eq:gvsg0}
\ee
where $g=4\pi\hbar^2 a/m$ is the effective coupling constant
and $\epsilon_{\mathbf{k}}$ the energy of a single particle 
with wavevector $\mathbf{k}$
on the lattice.
In fixed-node Monte~Carlo calculations, one rather uses an interaction potential in continuous real
space, e.g.\ a square well potential \cite{Pandha,Giorgini,Reddy,Lobo,Lobo_unb,Giorgini_unb}.

There is now a consensus about the fact that,
in the zero-range limit,
there
is a convergence of predictions for the two models $H_1$ and $H_2$,
in particular for the equation of state of the gas. 
The zero-range limit means, for single channel models, that $k_F b\ll 1$,
$b\ll \lambda$, and $b\ll |a|$, where $k_F$ is the Fermi wavevector and $\lambda$ is the
thermal de Broglie wavelength.\footnote{
The assumption $b\ll |a|$ is not necessary on the BCS side if one restricts to macroscopic observables such as the total energy, but it is necessary for microscopic observables such as the ones considered in section~\ref{subsec:srpc}.}
For two-channel models,
one has to impose the same conditions, not only for the interaction range $b$
but also for
the effective range $r_e$ (one indeed has $|r_e|\gg b$ on narrow Feshbach resonances \cite{PetrovBosons,YvanVarenna});
we shall see in appendix~\ref{app:2body} that one also needs to impose 
a condition on the background scattering length,
and that all conditions are satisfied in the experiment \cite{Hulet}.\footnote{
A simple formal definition of
the zero-range limit
for the homogeneous gas
 is to take the limit of vanishing density $k_F\to0$
while tuning the magnetic field $B$ in such a way that $1/(k_F a)$ remains constant,
and keeping constant all parameters of the Hamiltonian other than $B$
;
one then also has to take the limit
$T\to0$ so that $T/T_F$ remains constant.}

One can however consider observables which exist in reality
and are included in the two-channel model, but
which are absent in a single channel model. The most natural example
is the number of closed-channel molecules, which was recently measured 
at Rice using laser molecular excitation techniques
\cite{Hulet}.

In this paper, we show that the equilibrium 
number of closed-channel molecules $N_b$  
is directly
related to the equation of state of the gas, more precisely
to the derivative of the gas energy (or free energy at non-zero
temperature) with respect to $1/a$.  Paradoxically,
we can thus use the equation of
state calculated in \cite{Giorgini} within a single channel model,
in order
to predict $N_b$ in the crossover and compare to Rice measurements, see
section \ref{sec:prediction}.
Since the derivative of the energy with respect to $1/a$ is related to other
observables involving atomic properties only \cite{Tan_nkREVTEX,Tan_dEREVTEX}, it is also possible in principle to
access $N_b$ by a pure atomic measurement, see section \ref{sec:autres}.
We conclude in section~\ref{sec:conclusion}.

\section{Prediction for the number of closed-channel molecules}
\label{sec:prediction}

\subsection{General result}
We first suppose that the system is in an arbitrary eigenstate $|\psi\rangle$ 
of the two-channel
Hamiltonian $H_2$, with eigenergy $E$. From Hellmann-Feynman theorem
we have
\be
\frac{dE}{dB} = \langle \psi| \frac{dH_2}{dB}|\psi\rangle.
\ee
The only magnetic field dependent part of the Hamiltonian $H_2$ is the
bare closed-channel molecule energy $E_b(B)$ so that
\be
\frac{dE}{dB} = \mu_b N_b
\label{eq:dEdB}
\ee
where 
\be
N_b = \int d^3r \langle \psi_b^\dagger(\rr) \psi_b(\rr)\rangle
\ee
is the mean number of closed-channel molecules. 

We then eliminate the magnetic field by using as a parameter the scattering
length $a$ experienced by the trapped atoms
rather than $B$. In the general case where the trapping potential
$U_b(\mathbf{r})$ for the closed-channel molecule differs from twice 
the trapping potential $U(\mathbf{r})$ experienced by an open-channel atom, the
scattering length of two atoms around point $\mathbf{r}$ 
depends on $\mathbf{r}$, see appendix \ref{app:Beff}. To lift the ambiguity
we thus take for $a$ the scattering length in the trap center taken as the 
origin of coordinates, $\mathbf{r}=\mathbf{0}$.
We then rewrite (\ref{eq:dEdB}) as
\be
N_b =\frac{dE}{d(1/a)} \, \frac{d(1/a)}{dB} \, \frac{1}{\mu_b}.
\label{eq:nb}
\ee
Assuming that $\mu_b$ is magnetic field independent, $a$ as a function of $B$
can be calculated explicitly for $H_2$ by solving the
zero-energy two-body scattering problem in free space (see appendix~\ref{app:2body}),
and by including the energy shifts due to the trapping potentials
$U_b(\mathbf{0})$ and $U(\mathbf{0})$, as explained in the appendix
\ref{app:Beff}. The function takes the form
\be
a=a_{\rm fs}(B-\delta B_0)
\label{eq:acfb}
\ee
where we have introduced the function giving the free space scattering length
as a function of $B$,
\be
a_{\rm fs}(B) = a_{\rm bg} \left(1- \frac{\Delta B}{B-B_0}\right).
\label{eq:afs}
\ee
Here $a_{\rm bg}$ is the background scattering length, $B_0$ is the location
of the Feshbach resonance in the absence of trapping potential,
$\delta B_0=[2U(\mathbf{0})-U_b(\mathbf{0})]/\mu_b$ the shift of the resonance
location due to the trapping potential, and $\Delta B$ the resonance width.

Equation (\ref{eq:nb}) is directly applicable at zero temperature.
At non-zero temperature, one has to take a thermal average of (\ref{eq:nb}),
keeping in mind that the mean value of a derivative is not necessarily
equal to the derivative of the mean value. In the canonical ensemble
one can check the exact relation \cite{Bogoliubov_livre}
\be
\left\langle \frac{dE}{d(1/a)}\right\rangle = 
\left(
\frac{d F}{d(1/a)}
\right)_T
=
\left(\frac{d\langle E\rangle}{d(1/a)}\right)_S
\label{eq:Bogo}
\ee 
where $\langle \ldots\rangle$ is the thermal average,
the derivative of  the free energy $F$ is taken for a fixed temperature $T$,
and the derivative of $\langle E\rangle$  is taken for a fixed entropy $S$.
 We thus have in the canonical ensemble
\be
\langle N_b\rangle =\left(\frac{d\langle E\rangle}{d(1/a)}\right)_S \, \frac{d(1/a)}{dB} \, \frac{1}{\mu_b}.
\label{eq:Nb_Tfinie}
\ee

\subsection{Analytical results in a trap in limiting cases}
\label{subsec:ariatilc}

We now restrict to a spin balanced gas where the 
number of particles is equal to $N/2$ in each spin component. 
We also assume that the gas is at zero temperature 
(or at temperatures much smaller than the Fermi temperature $T_F$),
in a harmonic trap
\be
U(\rr) =  U({\bf 0}) 
+ \frac{1}{2} m \sum_{\alpha=x,y,z} \omega_\alpha^2 r_\alpha^2,
\label{eq:Uharm}
\ee
in the macroscopic regime where $k_B T_F$ 
is much larger than the oscillation quanta $\hbar\omega_\alpha$.
The usual definition of the Fermi temperature $T_F$ and of the Fermi
wavevector $k_F$ is then
\be
 k_B T_F = \frac{\hbar^2k_F^2}{2m} = (3N)^{1/3}  \hbar \bar{\omega}
\label{eq:ef}
 \ee
where $\bar{\omega}$ is the geometric mean of the 
three oscillation frequencies $\omega_\alpha$.
In this macroscopic regime, 
we rely on the local density approximation
to calculate $dE/d(1/a)$ for the trapped gas: This is expected to
be exact in the large $N$ limit, and to already hold
for the Rice experiment \cite{Hulet} since
$k_B T_F / (\hbar \omega_\alpha) \gtrsim 9$.

The key ingredient of the local density approximation
is the equation of state of the interacting homogeneous gas.
At zero temperature, and
in the zero-range limit detailed in the introduction, 
the equation of state is expected to be a universal function of the gas
density and of the scattering length, independent of the microscopic details 
of the interaction. Furthermore,
the spatial dependence of the scattering length in presence of the
trapping potential, due
to $U_b(\mathbf{r}) \neq 2\, U(\mathbf{r})$, is expected to
be negligible in the zero-range limit, 
as argued in the appendix~\ref{app:Beff}.
Under these assumptions, the local-density approximation for the
trapped gas leads to the universal form
\be
-\frac{dE}{d(1/a)} = 
\frac{\hbar^2 k_F}{m}\, N \, \mathcal{F} \left(\frac{1}{k_F a}\right),
\label{eq:genf}
 \ee
where $\mathcal{F}$ is a dimensionless function, expressed in the appendix
\ref{app:LDA} in terms of the dimensionless universal function $f$ 
giving the energy per particle $\epsilon^{\rm hom}$ of the homogeneous
interacting gas:
\be
\epsilon^{\rm hom}=\epsilon_0^{\rm hom} f\left(\frac{1}{k_F^{\rm hom} a}\right).
\label{eq:def_f}
\ee
Here $\epsilon_0^{\rm hom}$ and $k_F^{\rm hom}$ are the energy per particle
and the Fermi wavevector of the ideal Fermi gas with the same total density $n$
as the interacting gas:
\bea
k_F^{\rm hom} &=& (3\pi^2 n)^{1/3} \label{eq:defkhom}\\
 \epsilon_0^{\rm hom} &=& \frac{3}{10} \frac{\hbar^2 \left(k_F^{\rm hom}\right)^2}{m}.
\eea
 
In the BEC regime $0< k_F a \ll 1$ we have~\cite{LeyronasLHY,HuangYang,LeeYang}
\bea
f\left(\frac{1}{k_F^{\rm hom}a}\right)&=&-\frac{5}{3(k_F^{\rm hom}a)^2}
\label{eq:serie_hom_bec}
\\ & + & \nonumber  \frac{5}{18\pi}k_F^{\rm hom}a_d\left[
1+\frac{128}{15\sqrt{6\pi^3}}(k_F^{\rm hom}a_d)^{3/2}
\right]+\ldots
\eea
 where the first term comes from the dimer binding energy $\hbar^2/(ma^2)$, the second term is the mean field  interaction energy among the dimers, proportional to the dimer-dimer scattering
 length $a_d$,
 and the last term is the (bosonic) Lee-Huang-Yang correction.
 The solution of the $4$-fermion problem gives \cite{PetrovPRL,Leyronas}
 \be
 a_d\simeq 0.60 a.
\label{eq:ad}
 \ee
 The local-density approximation then leads to
 \be
 \mathcal{F}\left(\frac{1}{k_F a}\right) = \frac{1}{k_F a} 
 +\frac{5^{2/5}}{2^{12/5} \cdot 7} \left(\frac{a_d}{a}\right)^{2/5} (k_F a)^{7/5}+\ldots
\label{eq:serie_bec}
 \ee
 where we have omitted the Lee-Huang-Yang correction.
 
 In the BCS limit $0 < - k_F a \ll 1$,
 we have~\cite{HuangYang,LeeYang,Abrikosov,Galitski}
 \bea
 f\left(\frac{1}{k_F^{\rm hom}a}\right)&=&1+\frac{10}{9\pi} k_F^{\rm hom} a
 \nonumber
 \\&+&\frac{4(11-2\ln 2)}{21\pi^2} (k_F^{\rm hom}a)^2+\ldots
 \label{eq:serie_hom_bcs}
 \eea
 where the first term is the ideal gas result, the second term is the Hartree-Fock mean field term, and the last term is the (fermionic) Lee-Huang-Yang correction. This leads to~\footnote{
For a small {\it positive} scattering length $0<k_F a \ll 1$,
Eqs.(\ref{eq:serie_hom_bcs},\ref{eq:serie_bcs}) also apply to the {\it atomic} gas state;
since
the interaction potentials considered here have a two-body bound state for $a>0$,
this atomic gas state is only metastable~\cite{LudoYvanToyModel},
the ground state being the BEC of dimers considered previously [cf. Eq.(\ref{eq:serie_bec})].}
 \begin{multline}
\mathcal{F}\left(\frac{1}{k_F a}\right)=
\frac{512}{945 \pi^2} (k_F a)^2 \Big[
1 + \\
\left(\frac{256}{35\pi^2}-\frac{63+189\ln 2}{1024}\right) k_F a +\ldots
\Big].
 \label{eq:serie_bcs}
\end{multline}

Around the unitary limit $1/(k_F a)\to 0$ we use the expansion
 \be
 f\left(\frac{1}{k_F^{\rm hom} a}\right) = 
 \xi -\frac{\zeta}{k_F^{\rm hom} a}+\ldots,
\label{eq:expan}
 \ee
which gives for the trapped gas:
 \be
 \mathcal{F}(0) = \frac{2^7}{5^2\cdot 7 \cdot \pi} \frac{\zeta}{\xi^{1/4}} \simeq 0.28
\label{eq:Funi}
 \ee
 where we took the estimates 
 \bea
 \xi &\simeq & 0.41 
 \label{eq:valeur_ksi}
 \\
 \zeta &\simeq & 0.95\ .
\label{eq:valzeta}
 \eea
 The estimate for $\xi$ is obtained by averaging the predictions of
 \cite{Giorgini} ($\xi=0.42(1)$), of \cite{Juillet} ($\xi=0.449(9)$), of \cite{Reddy} ($\xi=0.42(1)$)
 and of~\cite{BulgacCrossover} ($\xi=0.37(5)$). The estimate for $\zeta$
 is obtained from the calculation of the short range pair correlation
 of  \cite{Lobo}, see subsection \ref{subsec:srpc}, and is close
 to the value $\zeta\simeq 1.0$ that may be extracted directly from the values of $f$ in
 \cite{Giorgini} by approximating the derivative by a two-point
 formula.
 \footnote{The values (\ref{eq:valeur_ksi},\ref{eq:valzeta}) of $\xi$ and $\zeta$
 can be inserted into the expressions~\cite{BulgacModes,CombescotLeyronasComment} for the derivatives of the frequencies of the hydrodynamic collective modes with respect to $1/(k_F a)$ taken at unitarity.
 The resulting slope 
 is compatible with the experimental data for the radial mode~\cite{GrimmModes,ThomasModePRA}, while
 the agreement with the data for the axial mode~\cite{GrimmModeVieux} is only marginal.
 The data from~\cite{GrimmModes} agree very well 
 for all positive values of $a$
 with the theoretical result~\cite{AstrakharchikMode} obtained from the fixed-node Monte Carlo equation of state~\cite{Giorgini}, while in~\cite{ThomasModePRA,GrimmModeVieux} finite-temperature effects may play a role~\cite{GrimmModes,AstrakharchikMode,Urban}.
}

%
\begin{figure}[h]
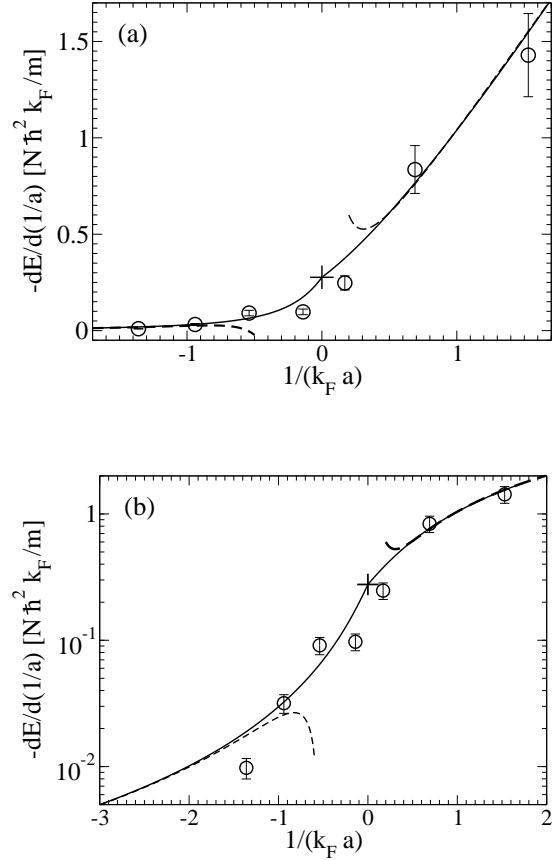

\begin{center}
\vskip 1cm
\resizebox{0.4\textwidth}{!}{%
\includegraphics{F.eps}}
\vskip 1cm
\resizebox{0.4\textwidth}{!}{%
\includegraphics{Flog.eps}}
\end{center}
\caption{Derivative 
\label{fig:F}       
$-dE/d(1/a)$ of the energy $E$ of a trapped two-component
Fermi gas, where $a$ is the $s$-wave scattering length.
Solid line: Theoretical prediction combining the Lee-Huang-Yang formulas (\ref{eq:serie_hom_bec},\ref{eq:serie_hom_bcs})
in the weakly interacting regimes with an interpolation of the fixed-node
Monte~Carlo results of \cite{Giorgini,Lobo} in the strongly interacting regime
(see appendix \ref{app:spline}).
Dashed lines: Analytical predictions in the weakly interacting
regimes, on the BEC side (\ref{eq:serie_bec})
and on the BCS side (\ref{eq:serie_bcs}).
Cross: Result (\ref{eq:Funi}) at unitarity.
Circles (with error bars): Experimental measurement of the number $N_b$
of closed-channel molecules in a lithium gas at Rice 
\cite{Hulet}, combined with the present theory linking $dE/d(1/a)$ to $N_b$,
see (\ref{eq:nb_jolie}).
(a): Linear scale on the vertical axis. (b): Logarithmic scale.
$-dE/d(1/a)$ is expressed in units of $N \hbar^2 k_F/m$.
All the theoretical predictions are obtained in the local-density approximation.
}
\end{figure}

\subsection{Comparison to Rice experiment}
\label{subsec:rice}

At Rice \cite{Hulet} the quantity $Z=N_b/(N/2)$ was measured for a lithium gas, 
by
resonantly exciting the closed-channel molecules with a laser  that 
transfers them to another, short-lived molecular state; the 
molecule depletion reflects into a reduction of the atom
number that can be measured.

We now compare the Rice results to our prediction (\ref{eq:nb}).
We first insert (\ref{eq:acfb},\ref{eq:afs}) into (\ref{eq:nb}) which gives
\be
N_b = N k_F R_* \mathcal{F}\left(\frac{1}{k_F a}\right) 
\left(1-\frac{a_{\rm bg}}{a}\right)^2
\label{eq:nb_jolie}
\ee
where the function $\mathcal{F}$ is defined in (\ref{eq:genf}).
The length $R_*$ depends on the interchannel coupling and is related
to the width of the Feshbach resonance as:
\be
\label{eq:defR*}
R_* = \frac{\hbar^2}{m a_{\rm bg} \mu_b \Delta B}.
\ee
For the  $B_0\simeq 834$G Feshbach resonance in $^6{\rm Li}$, we have $\mu_b\simeq2\mu_B$ where $\mu_B$ is the Bohr
magneton \cite{Koehler_review}, $\Delta B=-300$G, $a_{\rm bg}=-1405 a_0$ where $a_0=0.0529177$~nm is the Bohr radius
\cite{InnsbruckLi}; this gives $R_*=0.0269$~nm.
From the Rice data for $N_b/N$, using (\ref{eq:nb_jolie}),
we thus get values of $\mathcal{F}$, i.e. of $dE/d(1/a)$, 
that we compare in Fig.\ref{fig:F} to 
theoretical predictions, given in the local-density approximation
either by the expansions (\ref{eq:serie_bec},\ref{eq:serie_bcs}) 
in the weakly interacting
regime, or by (\ref{eq:Funi}) at the unitary limit,
or by an interpolation of the Monte~Carlo 
results of \cite{Giorgini,Lobo} in the strongly interacting regime
(see appendix \ref{app:spline} for details)
\footnote{
The error bars in Fig. 1 include the systematic theoretical uncertainty resulting from the leading order correction to (\ref{eq:afs}) given in \cite{InnsbruckLi}, but are largely dominated by
the experimental error bars from \cite{Hulet} because we did not include the three experimental points 
from \cite{Hulet}
which are deep in the BEC regime [$1/(k_F a)>5$];
in this regime, two-body theories which go beyond our two-channel model give good
 agreement with the Rice measurements~\cite{Hulet,Koehler_review,StoofZ}.
 }
.
The agreement is satisfactory, the experimental points being scattered
around the theoretical prediction. In particular, a possible shift $\delta B_0$
of the resonance location due to the fact that  $U_b(\mathbf{r}) \neq 2 U(\mathbf{r})$,  
see (\ref{eq:acfb}), an effect not explicitly included in \cite{Hulet} for the
calculation of $1/(k_F a)$ as a function of $B$, and which would manifest itself
in horizontal shifts of the experimental data with respect to theory,
does not show up in Fig.\ref{fig:F}. 
For the considered broad resonance of lithium, this is not
surprising, see appendix \ref{app:Beff}.

Our approach even allows to predict the time evolution
of the particle number, in presence of the depleting laser,
under the assumption that the molecular depletion rate is so weak 
that the equilibrium relation (\ref{eq:nb_jolie}) holds at all times.
Neglecting the slow off-resonant free-bound photoassociation process~\cite{Hulet}, we have
\be
\frac{dN}{dt} = - 2 N_b(t) \frac{\Omega^2}{\gamma}
\label{eq:cin}
\ee
where $\Omega^2/\gamma$ is the effective decay rate of the
closed-channel molecule,
the Rabi frequency $\Omega$ and the spontaneous emission
rate $\gamma$ being defined in \cite{Hulet}.\footnote{Eq. (\ref{eq:cin}) can be obtained from the
following master equation for the density matrix $\hat{\rho}$ of the gas:
$d\hat{\rho}/dt = (\Omega^2 / \gamma) \int d^3r \left[\psi_b(\mathbf{r}) \hat{\rho} 
\psi_b^\dagger(\mathbf{r})-\frac{1}{2} \{\psi_b^\dagger(\mathbf{r}) \psi_b(\mathbf{r}),\hat{\rho}\}\right]
+  [H_2,\hat{\rho}]/ (i\hbar)$.
$N$ in (\ref{eq:cin}) stands for $N_{\rm f} + 2 N_b$, where $N_{\rm f}$ is the
number of fermions;
thus $N\simeq N_{\rm f}$ in the regime considered in this paper.}

Simple expressions can be obtained in limiting cases:
In the BEC limit, the dominant term of the expansion (\ref{eq:serie_bec}) gives
\be
N(t)=N(0)\,e^{-\Gamma_0 t};
\ee
in the unitary limit, (\ref{eq:Funi}) gives
\be
N(t)=\frac{N(0)}{\left(1+\Gamma_0 t/6\right)^6};
\label{eq:N(t)_uni}
\ee
and in the BCS limit
the dominant term of (\ref{eq:serie_bcs}) gives
\be
N(t)=\frac{N(0)}{\left(1+\Gamma_0 t/2\right)^2}.
\label{eq:N(t)_BCS}
\ee
Here, $\Gamma_0 = -[dN(0)/dt]/N(0)$
is the inital atom loss rate, which can be expressed
in terms of the initial atom number $N(0)$ using 
Eqs.~(\ref{eq:nb_jolie},\ref{eq:cin}).
We see that in general $N(t)$ is not an exponential function
of time.
For the last point on the BCS side in \cite{Hulet},
the (unpublished) data for $N(t)$ are better fitted by
Eq.(\ref{eq:N(t)_BCS}) than by an exponential;
however the value of $\Gamma_0$ resulting from this new fitting procedure does not differ significantly from the published value \cite{HuletPrivate}.


\section{Related observables}
\label{sec:autres}
In this section, contrarily to subsections 2.2 and 2.3, we return
to the general case and we do not assume that the 
temperature is $\ll T_F$ and that the
numbers of atoms
in each spin component are equal.
For concreteness we take 
the lattice model (\ref{eq:H1}), even if the reasonings may be generalized to
 a continuous space model and to a two-channel model.

\subsection{Short range pair correlations}
\label{subsec:srpc}

As was shown in \cite{Tan_nkREVTEX,Tan_dEREVTEX} and recently rederived in \cite{Braaten},
in the zero interaction range limit,
the derivative of the gas energy with respect to the inverse
scattering length can be expressed in terms of the short range behavior
of the pair distribution function of opposite spin particles:
\be
-\left(\frac{d\langle E\rangle}{d(1/a)}\right)_S = \frac{4\pi\hbar^2}{m} \int d^3\mathbf{R}\,
\underset{r\to 0}{\tilde{\lim}} r^2 g_{\uparrow\downarrow}(\mathbf{R}+\rr/2,\mathbf{R}-\rr/2)
\label{eq:tan1}
\ee
where the pair distribution function is given by
\be
g_{\uparrow\downarrow}(\rr_1,\rr_2) =
\langle 
\psi_\uparrow^\dagger(\rr_1) 
\psi_\downarrow^\dagger(\rr_2)
\psi_\downarrow(\rr_2)
\psi_\uparrow(\rr_1) 
\rangle.
\ee
Since we are dealing here with a finite range interaction model,
we have taken a zero-range limit in (\ref{eq:tan1}), introducing
the notation
\be
\underset{r\to 0}{\tilde{\lim}} = \lim_{r\to 0} \lim_{b\to 0}.
\ee
Note that the order of the limits is important.
The limit in (\ref{eq:tan1}) is reached for a distance $r$ much smaller than 
$|a|$, the mean distance between particles,
and the thermal de~Broglie wavelength $\lambda$,
 but still much
larger than the interaction range $b$.

We thus see that a measurement of the pair distribution function
of the atoms gives access, using (\ref{eq:nb}), to the number
of closed-channel molecules.
Experimentally, the pair distribution function
was measured {\sl via}  the photoassociation rate in
a 1D Bose gas \cite{Gangardt,Weiss},
or simply by dropping the gas on a detector with high enough
spatio-temporal resolution \cite{Shimizu,Aspect_bosons,Aspect_fermions}.
The possibility of measuring the pair distribution function in the BEC-BCS crossover was
studied in \cite{Lobo,LoboScaling}.
In the unitary limit, by inserting into
Eq.(\ref{eq:tan1}) the value of $\tilde{\lim}\, r^2 g_{\uparrow\downarrow}$
calculated in \cite{Lobo} with the fixed-node Monte~Carlo technique, we get the value
(\ref{eq:valzeta}) for the parameter $\zeta$ defined in (\ref{eq:expan}).

The relation (\ref{eq:tan1}) is the three-dimensional version
of the relation obtained in \cite{Lieb} for a one-dimensional
Bose gas with contact interaction, using the Hellmann-Feynman theorem.
We now show that the Hellmann-Feyn\-man technique also provides a
simple derivation of (\ref{eq:tan1}). 
Taking the derivative of the gas eigenergy $E$ with respect 
to the effective coupling constant $g=4\pi\hbar^2 a/m$, for a fixed
lattice spacing $b$, the Hellmann-Feynman theorem gives
\bea
\label{eq:HFl1}
\left(\frac{d\langle E\rangle}{dg}\right)_S &=& \frac{dg_0}{dg}\left\langle \sum_\rr b^3
\psi_\uparrow^\dagger\psi_\downarrow^\dagger\psi_\downarrow\psi_\uparrow\right\rangle
\\
&=& \frac{dg_0}{dg}  \sum_\rr b^3 g_{\uparrow\downarrow}(\rr,\rr),
\label{eq:HFl2}
\eea
where $g_0$ is a function of $g$ as given in (\ref{eq:gvsg0}),
and where we used the relation $\langle dE/dg \rangle = (d\langle E\rangle/dg)_S$ [cp.~(\ref{eq:Bogo})].
In a gas, with an interaction range much smaller than the mean distance
$\sim 1/k_F$ between particles and than the thermal de~Broglie wavelength $\lambda$, 
the pair distribution function is dominated by two-body physics
if $|\rr_1-\rr_2|$ is much smaller than $1/k_F$ and $\lambda$
(and also much smaller than $a$ in the BEC regime), a
property contained in the Yvon \cite{Yvon} and Waldmann-Snider \cite{Snider}
ansatz of kinetic theory, which also appeared
 in the context 
of quantum gases \cite{Lobo,Huang_livre,Holzmann,Leggett_IHP}:
\be
g_{\uparrow\downarrow}(\rr_1,\rr_2) \simeq
C\left(\frac{\rr_1+\rr_2}{2}\right) \, |\phi(\rr_1-\rr_2)|^2
\ee
where $\phi$ is the two-body zero-energy free space scattering
state for two atoms, here
in the spin singlet state.
With the lattice model we find
\footnote{
An elegant derivation is to enclose two atoms
in a fictitious cubic box with periodic boundary conditions
of volume $L^3$, with $L\gg |a|,b$. To leading order in $a/L$,
the wavefunction is $\phi_{\rm box}(\rr_1-\rr_2)=\phi(\rr_1-\rr_2)/L^{3/2}$
with an energy given by the ``mean-field" shift
$E_{\rm box}= g/L^3$. The Hellmann-Feyn\-man theorem
(\ref{eq:HFl1}) then gives the result (\ref{eq:phi20}).
An alternative derivation is to directly calculate $\phi$ in Fourier space [see
(\ref{eq:ms})] which
gives $\phi(\rr)=1-g_0 \phi(\mathbf{0}) \int_{[-\pi/b,\pi/b]^3}
[d^3\kk/(2\pi)^3] \exp(i\kk\cdot\rr)/(2\epsilon_{\kk})$;
this, combined with (\ref{eq:gvsg0}), gives 
(\ref{eq:phi20}).
}
\be
\label{eq:phi20}
|\phi(\mathbf{0})|^2 = \frac{dg}{dg_0},
\ee
so that Eq.(\ref{eq:HFl2}) simplifies to
\be
\left(\frac{d\langle E\rangle}{dg}\right)_S \simeq \sum_\rr b^3 C(\rr).
\label{eq:HFl3}
\ee
On the other hand, 
at interparticle distances much larger than $b$, the zero-energy
scattering wavefunction $\phi$ behaves as $1-a/|\rr_1-\rr_2|$ so that
\be
\underset{r\to 0}{\tilde{\lim}} r^2 g_{\uparrow\downarrow}(\mathbf{R}+\rr/2,\mathbf{R}-\rr/2)
= a^2 C(\mathbf{R}).
\ee
Eq.(\ref{eq:HFl3}) thus leads to (\ref{eq:tan1}).

\subsection{Tail of the momentum distribution}

As was shown in \cite{Tan_dEREVTEX} and recently rederived in \cite{Braaten},
in the zero-range limit,
the momentum distribution of the gas has a large momentum tail
scaling as $1/k^4$, with a coefficient common to the two spin states
and proportional to $-dE/d(1/a)$:
\be
\underset{k\to+\infty}{\tilde{\lim}} k^4 n_\sigma(k) = -\frac{4\pi m}{\hbar^2} 
\left(\frac{d\langle E\rangle}{d(1/a)} \right)_{\!S}
\label{eq:tan2}
\ee
where
\be
\underset{k\to +\infty}{\tilde{\lim}} = \underset{k\to +\infty}{\lim}\  \underset{b\to 0}{\lim},
\ee
$\sigma= \uparrow$ or $\downarrow$ and the momentum distribution is normalized
as
\be
\int \frac{d^3k}{(2\pi)^3} n_\sigma(\kk) = \langle N_\sigma\rangle,
\ee
where $\langle N_\sigma\rangle$ is the mean number of particle in the spin state $\sigma$.
When combined with (\ref{eq:nb}), this relation reveals that the
number of closed-channel molecules may be deduced from a 
measurement of the momentum distribution of the atoms.

The momentum distribution in a two-component Fermi gas
was actually already measured \cite{JILAnk,TarruellVarenna,JILA_T_finie}, simply
by switching off the interaction and measuring the cloud density
after a ballistic expansion.
Two practical problems however arise, the effect of the non-zero
switch-off time of the interaction \cite{JILAnk,Holland_nk} and the signal to noise
ratio in the far tails of the distribution;
we have therefore not been able to extract the value of $dE/d(1/a)$ from
the published data for $n_\sigma(k)$.
The value of $dE/d(1/a)$ has not been extracted either from the fixed-node Monte~Carlo calculations of $n_\sigma(k)$ performed in \cite{giorgini_nk}.

The relation (\ref{eq:tan2}) is a three-dimensional version
of the relation derived for a one-dimensional Bose gas in
\cite{Olshanii}.
We now present a very simple rederivation of (\ref{eq:tan2}).

In a gas with short range interactions, i.e.\
$b \ll 1/k_F$ and $b\ll \lambda$
(and $b\ll a$ on the BEC side),  the tail of the momentum
distribution is dominated by binary collisions in the spin
singlet channel between
a particle with a large momentum $\kk$ and a particle with a 
momentum $\simeq -\kk$, so that at large $k$
\be
n_{\sigma}(\kk) \simeq   \mathcal{B} |\tilde{\phi}(\kk)|^2
\label{eq:withB}
\ee
where $\mathcal{B}$ is momentum-independent
and $\tilde{\phi}(\kk)$ is the Fourier transform
$\sum_\rr b^3
e^{-i\kk\cdot\rr}\phi(\rr)$ of the zero-energy
two-body free space scattering state $\phi(\rr)$.
For the contact interaction in the lattice model
the two-particle Schr\"odinger's equation takes the form
\cite{boite}
\be
2\epsilon_{\kk} \tilde{\phi}(\kk) + g_0 \phi(\rr=\mathbf{0})=0,
\label{eq:ms}
\ee
where $\epsilon_\kk$ is the free wave dispersion relation in the lattice
model, so that $\tilde{\phi}(\kk)$ drops as $1/\epsilon_\kk$ for
large $k$, and the associated kinetic energy diverges for $b\to 0$.
Eq.(\ref{eq:withB}) implies that the kinetic energy of the gas
is proportional to the kinetic energy 
of the zero-energy two-body scattering state
in the zero-range limit:
\be
E_{\rm kin}[{\rm gas}] \simeq \mathcal{B} \cdot E_{\rm kin}[\phi].
\ee
Since the total energies have a finite limit for $b\to 0$, 
we can replace the kinetic energies by the opposite of the interaction
energies in the above formula:
\be
E_{\rm int}[{\rm gas}] \simeq \mathcal{B}\cdot E_{\rm int}[\phi].
\ee
The interaction energy
of the gas is $g_0 \left\langle \sum_\rr b^3
\psi_\uparrow^\dagger\psi_\downarrow^\dagger\psi_\downarrow\psi_\uparrow\right\rangle$,
equal to $(d\langle E\rangle/dg)_S\, g_0 (dg/d g_0)$
according to (\ref{eq:HFl1}).
The interaction energy of $\phi$ is simply $g_0 |\phi(\mathbf{0})|^2$,
equal to $g_0 (dg/d g_0)$ according to (\ref{eq:phi20}). 
The common factor $g_0 (dg/dg_0)$ simplifies and
\be
\left(\frac{d\langle E\rangle}{dg}\right)_S = \mathcal{B}.
\ee

It remains to relate the coefficient $\mathcal{B}$ 
to $\tilde{\lim}_{k\to+\infty} k^4 n_\sigma(k)$,
using (\ref{eq:withB}). In the zero-range limit,
the scattering state $\phi(\rr)$ tends to $1-a/r$, so that,
by Fourier transform,
$\tilde{\phi}(\kk)$ tends to $(2\pi)^3 \delta(\kk) -4\pi a/k^2$.
We then get (\ref{eq:tan2}).

\section{Conclusion}
\label{sec:conclusion}

We have calculated the number $N_b$ of closed-channel 
mole\-cu\-les in the whole BEC-BCS crossover for a two-component
Fermi gas.
Our result is in satisfactory agreement with the experimental results from Rice~\cite{Hulet}.
The expression (\ref{eq:nb}) for $N_b$ is 
proportional to the quantity $-dE/d(1/a)$ where
$E$ is the energy of the gas;
this quantity is universal
in the zero interaction range limit.
At unitarity we find $N_b/N=k_F R_* \mathcal{F}(0)$ where $\mathcal{F}(0)$ is a universal constant  given in (\ref{eq:Funi}) and the length $R_*$ is related to the resonance parameters by (\ref{eq:defR*});
in the zero range and zero effective range limit where $k_F b\ll1$ and $k_F |r_e|\ll1$ we have
$k_F R_*\ll1$ [see Eq.(\ref{eq:re_res})],
we thus confirm that
 $N_b\ll N$ i.e.
the gas mainly populates the open channel.
In the BCS limit, $-dE/d(1/a)$ and thus $N_b$ are determined by the Hartree-Fock mean field energy, and do not tend 
exponentially to zero with $1/(k_F|a|)$, 
contrarily to the prediction of usual BCS theory~\cite{Hulet}.
This is also related to the fact that the BCS theory does not predict the correct short-distance pair correlations~\cite{Iacopo}.

The quantity
$-dE/d(1/a)$ is 
 related
to the short-distance behavior of the pair correlation function and to the large-$k$ tail of the momentum distribution, as discovered by Tan \cite{Tan_nkREVTEX,Tan_dEREVTEX}
and rederived in \cite{Braaten} and in this paper.
The quantity $-dE/d(1/a)$ is also proportional to the average radiofrequency shift \cite{Zwerger,Baym}.
The general idea that various short-range two-body quantities such as the number of closed-channel molecules, the interaction energy of the gas and the average radiofrequency shift are the product
of the value of the considered observable in the zero-energy two-body scattering state
and a single universal many-body quantity was given by Leggett~\cite{Leggett_IHP}.
Our work shows that the Rice experiment \cite{Hulet} is the first direct measurement of this 
fundamental quantity in the BEC-BCS crossover.

It would be interesting to study
the effect of a
 finite temperature. Experimentally, this could be achieved by
measuring the
number of closed-channel molecules
together with the density profile of the trapped gas,
the released energy~\cite{Thomas1,SalomonCrossover},
or the entropy~\cite{JinPotentialEnergy,thomas_entropie}.
Theoretically,
 Eq.(\ref{eq:Nb_Tfinie}) can be used
when more Monte Carlo results at finite temperature will become available.
A useful tool in this context is the virial theorem:
at unitarity it directly gives the total energy in terms of the density profile~\cite{ThomasVirielExp,thomas_entropie,WernerSym,ThomasVirielTheorie} while for a finite $a$
it also involves $[d\langle E\rangle/d(1/a)]_S$~\cite{TanViriel,WernerViriel}.

\begin{acknowledgement}
We thank A.~Bulgac, F.~Chevy, M.~Colom\'e-Tatch\'e, R.~Hulet, M. Jona-Lasinio,
F.~Lalo\"e, P.~Naidon, S.~Nascimb\`ene, G.~Partridge, L. Pricoupenko and C.~Salomon
for very useful discussions and comments,
and G.~Partridge {\it et al.} for providing 
us with the data of \cite{Hulet} in numerical form.
Our group is a member of IFRAF.

\end{acknowledgement}

\noindent {\bf Note:} We became aware of the related work~\cite{BraatenLong}
while completing this paper,
and of~\cite{Zhang_Leggett}
while revising it.

\appendix

\section{Two-body analysis of the two-channel model in free space}
\label{app:2body}
\subsection{Scattering amplitude}

In this appendix we briefly discuss the two-body scattering properties for our model
Hamiltonian $H_2$ in free space, that is in the absence of trapping
potential ($U_b(\mathbf{r})\equiv 0, U(\mathbf{r})\equiv 0$). 
The scattering problem may be considered
in the center of mass frame so that we take the two-body scattering state
\begin{equation}
|\psi\rangle = \beta b_\mathbf{0}^\dagger |0\rangle
+\int\frac{d^3k}{(2\pi)^3} A(\mathbf{k})a_{\mathbf{k}\uparrow}^\dagger
a_{\mathbf{k}\downarrow}^\dagger |0\rangle,
\label{eq:appen_state}
\end{equation}
with the unknown amplitude $\beta$ in the closed channel
and the unknown momentum dependent amplitude $A(\mathbf{k})$
in the open channel.
The annihilation operators for the bosonic molecules,
$b_{\mathbf{k}}$, and for the fermionic atoms $a_{\mathbf{k}\sigma}$,
$\sigma=\uparrow$ or $\downarrow$, are the Fourier transforms
of the corresponding field operators $\psi_b(\mathbf{r})$
and $\psi_{\sigma}(\mathbf{r})$, where the Fourier transform of a
function $f(\mathbf{r})$ is $\tilde{f}(\mathbf{k})=
\int d^3r \exp(-i\mathbf{k}\cdot\mathbf{r}) f(\mathbf{r}).$
Inserting the state (\ref{eq:appen_state}) in Schr\"odinger's
equation $(H_2-E)|\psi\rangle=0$ 
and projecting respectively on the molecular subspace
and the atomic subspace leads to
\begin{eqnarray}
\label{eq:appen_mol}
(E_b-E) \beta + \Lambda \gamma = 0 \\
\left(\frac{\hbar^2k^2}{m}-E\right) A(\mathbf{k})
+\tilde{\chi}(\mathbf{k}) (\beta\Lambda + g_0 \gamma) =0
\label{eq:appen_at}
\end{eqnarray}
where we used the fact that the function $\chi(\mathbf{r})$ is real
and even, and where we introduced the amplitude
\begin{equation}
\label{eq:appen_gamma}
\gamma = \int\frac{d^3k}{(2\pi)^3} \tilde{\chi}(\mathbf{k})A(\mathbf{k}).
\end{equation}
We now specialize to a state corresponding to the
scattering of a spin $\uparrow$ atom with incoming wavevector $\mathbf{k}_0$
onto a spin $\downarrow$ atom with incoming wavevector $-\mathbf{k}_0$,
so that $E=\hbar^2 k_0^2/m$ is the energy of the incoming state.
Eliminating $\beta$ in terms of $\gamma$ with (\ref{eq:appen_mol})
and using (\ref{eq:appen_at}) leads to
\begin{equation}
\label{eq:appen_pf}
A(\mathbf{k}) = (2\pi)^3 \delta(\mathbf{k}-\mathbf{k_0})+
\frac{\gamma \tilde{\chi}(\mathbf{k})[g_0+\Lambda^2/(E-E_b)]}
{E+i0^+-\hbar^2 k^2/m}
\end{equation}
where we have singled out the incoming state, and
the scattered component is guaranteed to correspond
to a pure outgoing wave in real space by the introduction
of $+i0^+$ in the denominator. Injecting (\ref{eq:appen_pf})
in (\ref{eq:appen_gamma}) gives a closed linear equation for $\gamma$.
From the known large $r$ expression for the kinetic energy Green's function
we extract the scattering amplitude
\begin{equation}
f_{k_0} = \frac{-m\tilde{\chi}(\mathbf{k}_0)^2/(4\pi\hbar^2)}
{[g_0+\frac{\Lambda^2}{\frac{\hbar^2 k_0^2}{m}-E_b}]^{-1} - \int \frac{d^3k}{(2\pi)^3}
\frac{\tilde{\chi}^2(\mathbf{k})}{\frac{\hbar^2 k_0^2}{m}+i0^+-\frac{\hbar^2 k^2}{m}}}.
\label{eq:app_fk_chi_quelconque}
\end{equation}
Since we are considering a $s$-wave Feshbach resonance, 
the function $\chi(\mathbf{r})$ is rotationally invariant, a property
that we have used in (\ref{eq:app_fk_chi_quelconque}).

We now perform explicit calculations of various quantities characterizing
the low energy scattering, such as the scattering length and the
effective range. To this end, it is convenient as in
\cite{YvanVarenna} to perform the specific choice 
\begin{equation}
\tilde{\chi}(\mathbf{k}) = \exp(-k^2 b^2/2)
\end{equation}
and to extend (\ref{eq:app_fk_chi_quelconque}) to negative energies,
setting $k_0 = i q, q>0$ and $E=-\hbar^2 q^2/m$. One then obtains
an explicit expression 
\begin{equation}
\label{eq:appen_erf}
-\frac{1}{f_{iq}} =  e^{-q^2 b^2}
\left[\frac{4\pi\hbar^2/m} {g_0-\frac{\Lambda^2}{E_b+\hbar^2q^2/m}}
+\frac{1}{b\sqrt{\pi}}\right] + q\, \mathrm{erf}\,(qb)-q
\end{equation}
where the last term $-q$ is imposed by the optical theorem and
$\mathrm{erf}$ is the error function.

The inverse of the scattering length is obtained by taking
$q\to 0$ in (\ref{eq:appen_erf}):
\begin{equation}
\label{eq:appen_inva}
\frac{1}{a} = \frac{4\pi\hbar^2/m}{g_0-\Lambda^2/E_b} + \frac{1}{b\sqrt{\pi}}.
\end{equation}
The large $E_b$ limit of this expression is the inverse
of the background scattering length:
\begin{equation}
\frac{1}{a_{\rm bg}} = \frac{4\pi\hbar^2/m}{g_0} + \frac{1}{b\sqrt{\pi}}.
\end{equation}
The location of the Feshbach resonance corresponds
to an energy $E_b^0$ of the closed-channel molecule such that
$1/a=0$, so that
\begin{equation}
E_b^0 = \frac{\Lambda^2}{g_0 + 4\pi^{3/2} \hbar^2 b/m}.
\end{equation}
Supposing that the energy $E_b$ of the closed-channel molecule
varies linearly with the magnetic field $B$,
\begin{equation}
E_b(B) = E_b^0 + \mu_b (B-B_0),
\end{equation}
and introducing the (possibly negative)
resonance width $\Delta B$ such that
\begin{equation}
\mu_b \Delta B = \frac{\Lambda^2}{g_0} - E_b^0,
\end{equation}
we get from (\ref{eq:appen_inva}) exactly the $B$-field dependence
(\ref{eq:afs}) for the scattering length $a$ in free space.

We now consider the effective range $r_e$ defined by the low-$q$
expansion $-1/f_{iq} = a^{-1} - q + \frac{1}{2} q^2 r_e + O(q^4)$.
Its general expression for our model is
\begin{equation}
r_e = \frac{-8\pi\hbar^4\Lambda^2/m^2}{(g_0 E_b-\Lambda^2)^2}
+\frac{4b}{\sqrt{\pi}} -\frac{2b^2}{a}.
\end{equation}
Of particular interest is the value of the effective range right
on resonance,
\begin{equation}
r_e^{\rm res}= -2 R_* + \frac{4b}{\sqrt{\pi}},
\label{eq:re_res}
\end{equation}
where the length $R_*$ is always non-negative,
\begin{equation}
R_* = \left(\frac{\Lambda}{2\pi b E_b^0}\right)^2.
\end{equation}
This definition of $R_*$ agrees with Eq.~(\ref{eq:defR*}).

A very convenient rewriting of the scattering amplitude 
$f_{iq}$ is obtained by taking as independent parameters,
in addition to the van der Waals length $b$,
the two lengths $a_{\rm bg}, R_*$ 
rather than the bare parameters $g_0, \Lambda$ of the Hamiltonian.
The quantity $E_b-\Lambda^2/g_0$ appearing in (\ref{eq:appen_erf})
is conveniently rewritten using the quantity
\begin{eqnarray}
Q^2 &\equiv& \frac{m}{\hbar^2 g_0} \left(g_0 E_b-\Lambda^2\right) \\
&=& \frac{-1}{a_{\rm bg} R_*} \frac{1}{1-a_{\rm bg}/a}
\end{eqnarray}
which can be positive or negative.
The ratio $g_0/\Lambda$ is also easily eliminated with the identity
\begin{equation}
\frac{g_0^2}{4\pi\Lambda^2} = R_* a_{\rm bg}^2.
\end{equation}
Our final expression for the scattering amplitude is thus
\begin{equation}
\label{eq:appen_jolie}
-\frac{1}{f_{iq}} = \frac{e^{-q^2b^2}}{a} 
\left[1-\left(1-a/a_{\rm bg}\right)\frac{q^2}{q^2+Q^2}\right]
+q\, \mathrm{erf}\,(qb) - q
\end{equation}
i.e.
\begin{equation}
\label{eq:appen_fk}
-\frac{1}{f_{k}} = \frac{e^{k^2b^2}}{a} 
\left[1-\left(1-a/a_{\rm bg}\right)\frac{k^2}{k^2-Q^2}\right]
-i k\, \mathrm{erf}\,(-i k b) + i k.
\end{equation}
We deduce
the effective range in terms of $a, a_{\rm bg}, R_*$ and $b$:
\begin{equation}
r_e = -2 R_* (1-a_{\rm bg}/a)^2 +\frac{4b}{\sqrt{\pi}}-\frac{2b^2}{a}.
\label{eq:re_2_canaux}
\end{equation}

\subsection{Conditions for reaching the zero-range limit}

In a degenerate gas, we expect that the zero-range limit is reached and the equation of state $E(1/a)$ becomes model independent provided the conditions
\bea
k b &\ll& 1 \label{eq:kb<<1} \\
- \frac{1}{f_k} &\simeq& \frac{1}{a} + i k
\label{eq:fkPP}
\eea
hold for $k\sim k_F$ and for $k\sim 1/a$ on the BEC side \cite{YvanVarenna}.

We now assume that (\ref{eq:kb<<1}) holds, and we derive validity conditions
for (\ref{eq:fkPP})
in the case of our two-channel model.
Following the usual procedure, we use the second order expansion
\be
- \frac{1}{f_k} = \frac{1}{a} + i k - \frac{1}{2} r_e k^2 + \dots,
\label{eq:fk_re}
\ee
which reduces to (\ref{eq:fkPP}) if the effective range is small enough to have
\be
\frac{1}{2} |r_e| k^2 \ll \left| \frac{1}{a}+i k \right|.
\label{eq:kre}
\ee
Taking into account the condition (\ref{eq:kb<<1}) 
as well as the expression (\ref{eq:re_2_canaux}) of $r_e$ for the two-channel model, and 
using $|i+1/(ka)|\geq 1$
and $|1+ika|\geq 1$, we find that the condition (\ref{eq:kre})
becomes equivalent to
\be
k^2 R_* \left(1-\frac{a_{\rm bg}}{a}\right)^2 \left| \frac{1}{a}+i k \right|^{-1} \ll 1.
\label{eq:cond_re}
\ee
A second condition is obtained by imposing that the higher order terms $\dots$ in (\ref{eq:fk_re})
are indeed negligible compared to $k^2 |r_e|/2$. For this to hold we need not only (\ref{eq:kb<<1}), but also $k^2\ll |Q^2|$ i.e.
\be
k^2 |a_{\rm bg}| R_* \left| 1-\frac{a_{\rm bg}}{a}\right| \ll 1.
\label{eq:cond_abg}
\ee

We conclude that to reach the universal zero-range limit, it is sufficient to fulfill the three conditions (\ref{eq:kb<<1},\ref{eq:cond_re},\ref{eq:cond_abg}).
All three conditions are satisfied for the experimental data on $^6{\rm Li}$ from \cite{Hulet} reproduced in Fig.\ref{fig:F}, where $1/k_F>250$~nm and $|a|>200$~nm:
 using the values $|a_{\rm bg}|=74$~nm and $R_*=0.027$~nm (see section~\ref{subsec:rice})
 and taking the estimate $b\simeq 2.1$~nm obtained by imposing that the effective range on resonance (\ref{eq:re_res}) should be equal to the value $4.7$~nm obtained in the multi-channel calculation of~\cite{Strinati}, we find that the left hand sides of Eqs.(\ref{eq:kb<<1},\ref{eq:cond_re},\ref{eq:cond_abg}) are respectively smaller than $10^{-2}$, $3\cdot 10^{-4}$ and $7\cdot 10^{-5}$.

To be complete, let us point out that the conditions (\ref{eq:cond_re},\ref{eq:cond_abg})
are not stricly necessary since it is possible that the $\dots$  in (\ref{eq:fk_re})
are not negligible compared to $k^2 |r_e|/2$ and (\ref{eq:fkPP}) nevertheless holds.
We have chosen these conditions because they simplify the discussion, in particular 
they automatically ensure that the analytic continuation $f_{i q}$ has a pole at $q\simeq 1/a$ 
so that the dimer energy is $\simeq -\hbar^2/(m a^2)$.
For this last property,
in the regime $b\ll a$, one directly gets from (\ref{eq:appen_jolie})
the truly necessary condition
\be
\left|(1-a/a_{\rm bg})^{-1} +
(1-a_{\rm bg}/a)^{-2}
a/R_*\right|  \gg 1.
\ee
Close to the Feshbach resonance, $a\gg |a_{\rm bg}|$, this reduces to the simpler condition
\be
a\gg R_*.
\label{eq:simple}
\ee

\section{Effect of a difference between the trapping potentials seen by a closed-channel molecule and by an open-channel pair of atoms}
\label{app:Beff}

In this appendix, we consider the physical consequences of the fact that,
in general, the trapping potential $U_b(\rr)$ seen by a closed-channel molecule
is not simply the trapping potential $2\,U(\rr)$ seen by two separated atoms in the open channel:
\be
U_b(\rr)\neq 2 \, U(\rr),
\label{eq:sit}
\ee
and we derive conditions for the gas to behave approximately as if we had $U_b(\rr)=2\,U(\rr)$.
Note that a significant deviation in (\ref{eq:sit}) is not {\sl a priori} excluded
for the laser traps used e.g.\ in \cite{Hulet,InnsbruckLi} for lithium,
with a laser wavelength $\lambda_L\simeq 1\, \mu$m:
Contrarily to the atomic lightshift,
the lightshift of the closed-channel molecule may be sensitive to the laser detuning from
transitions to bound states of the $1 ^1\Sigma_u^+(A)$ 
Born-Oppenheimer interaction potential 
between a ground-state atom and an excited atom, considering the fact
that 
the minimum of
this excited-ground interaction potential is separated from the dissociation limit
of the ground-ground interaction potential by only $\simeq 0.6\, \mu{\rm m}^{-1}< 1/\lambda_L$ 
in spectroscopic units ~\cite{ChemPhys}.

\subsection{Position dependent scattering length and effective magnetic field}

At the two-body level, (\ref{eq:sit}) leads to an effective position dependent scattering
length for two opposite spin atoms. To define such a local scattering length
around point $\rr_0$, we consider the ``tangential'' free space problem
where the trapping potentials are replaced by the constant values 
$U_b(\rr_0)$ and $U(\rr_0)$ and we call $a(\rr_0)$ the resulting scattering
length. 
We simply have to revisit the calculations of appendix \ref{app:2body}:
In (\ref{eq:appen_mol}) one has to replace $E_b$ by $E_b + U_b(\rr_0)$ to include the trapping energy
shift of the closed-channel molecule; similarly in (\ref{eq:appen_at}) one has to replace 
$\frac{\hbar^2 k^2}{m}$ by $\frac{\hbar^2 k^2}{m}+2 U(\rr_0) $ to include the trapping
energy shift of the two atoms; finally, for atoms with incoming wavevectors $\pm\kk_0$,
the scattering state eigenenergy is now $E=\frac{\hbar^2 k_0^2}{m}+2 U(\rr_0)$.
The net effect on the scattering amplitude $f_{k_0}$ in (\ref{eq:app_fk_chi_quelconque})
is to apply the substitution
\be
E_b \longrightarrow E_b+U_b(\rr_0)-2 U(\rr_0),
\label{eq:subst}
\ee
so that $a(\rr_0)$ is simply obtained by applying (\ref{eq:subst}) to
(\ref{eq:appen_inva}).

A powerful interpretation of this substitution procedure may be given 
for a magnetic field independent $\mu_b$:
We can rewrite the term $E_b(B)+U_b(\rr)$ appearing in (\ref{eq:definition_H2}) as
$E_b[B_{\rm eff}(\rr)]+2\,U(\rr)$
provided we set
\be
B_{\rm eff}({\bf r})= B+
\frac{U_b(\rr)-2\,U(\rr)}{\mu_b}.
\label{eq:Beff(r)}
\ee
Thus we can replace $U_b(\rr)$ by $2\,U(\rr)$ in (\ref{eq:definition_H2})
provided we also replace the homogeneous magnetic field $B$ in (\ref{eq:definition_H2}) 
by the inhomogeneous effective magnetic field $B_{\rm eff}({\bf r})$.
One then immediately obtains the position dependence of the scattering length,
\be
a(\rr)=a_{\rm fs}[B_{\rm eff}(\rr)], 
\label{eq:a_de_r}
\ee
the function $a_{\rm fs}(B)$ being given in (\ref{eq:afs}).

This effective magnetic field picture is also very efficient at the many-body level.
In the grand canonical point of view, 
one adds the term $-\mu \hat{N}$ to the Hamiltonian $H_2$,
where $\mu$ is the chemical potential,
and the particle number operator $\hat{N}$ is defined as 
$\hat{N}=\hat{N}_f+2\hat{N}_b$, $\hat{N}_f$ giving the number of fermions and $\hat{N}_b$
the number of bosons, since this is the 
quantity which commutes with $H_2$.
In the local density approximation, in the general case $U_b\neq 2 U$,
we thus see that the gas behaves locally like a free space, homogeneous gas of chemical
potential $\mu-U(\rr)$, in presence of the locally homogeneous magnetic field $B_{\rm eff}({\bf r})$.

\subsection{Shift of the resonance location}

We now consider the case of an optical trap such as the one at Rice~\cite{Hulet}:
$U(\rr)$ and $U_b(\rr)$ are the lightshift potentials experienced 
by an atom and by a closed-channel molecule. Being both proportional to the laser intensity, they are mutually
proportional:
\be
U_b(\rr) = 2(1+\eta) U(\rr)
\ee
where the dimensionless parameter $\eta$ is a convenient measure of the relative deviation
between $U_b$ and $2U$.
At large $\rr$ the laser intensity vanishes, so do $U_b$ and $U$.

In practice, the gas is trapped around the intensity maximum of the laser field,
located in $\rr=\mathbf{0}$,
and the extension of the cloud along the axial (resp. radial) direction
is small compared to the Rayleigh length (resp. the waist) of the laser beam;
thus, to leading order, $U(\rr)\simeq U(\mathbf{0})$,
we can temporarily forget about the spatial variation of $B_{\rm eff}$ and 
take $B_{\rm eff}\simeq B-\delta B_0$,
where
\be
\delta B_0 = \frac{-U_b(\mathbf{0}) + 2\, U(\mathbf{0})  }{\mu_b} 
= -\frac{2\eta U(\mathbf{0})}{\mu_b}.
\ee
Since the free space scattering length $a_{\rm fs}(B)$ only depends on the difference $B-B_0$, 
replacing $B$ by $B_{\rm eff}$ is equivalent to shifting the position $B_0$ 
of the Feshbach resonance by the quantity $\delta B_0$.
Introducing the length $R_*$ defined in  (\ref{eq:defR*}),
the ratio of this shift to the width $\Delta B$ of the Feshbach resonance may be written as
\be
\frac{\delta B_0}{\Delta B}=-\frac{2 m U(\mathbf{0})}{\hbar^2} a_{\rm bg} R_* \eta.
\label{eq:dB0surDB0}
\ee
In the Rice experiment, the trap depth $-U(\mathbf{0})$ is at most 
$k_B\times2.71~\mu{\rm K}$~\cite{HuletPrivateData}, so that we have
\be
\frac{|\delta B_0|}{|\Delta B|} < \frac{|\eta|}{7500}
\label{eq:dB0<<DeltaB}
\ee
where we have used the values of $a_{\rm bg}$ and $R_*$ given below (\ref{eq:defR*}) and
the mass of lithium. Except if $\eta$ takes truly large values for the 
laser wavelength $\lambda_L=1064$~nm used in \cite{Hulet}, we expect only a small 
shift of the resonance location due to the laser trap, relative to the resonance width.

In \cite{Hulet} the value of the scattering length as a function of $B$ was calculated essentially from
the free space formula $a_{\rm fs}(B)$ ignoring the shift $\delta B_0$ due to the trap, and
taking for the resonance location $B_0$ the value measured in \cite{InnsbruckLi} in a laser trap with 
different laser wavelength $\lambda_L=1030$~nm and intensity. On the contrary, our definition
of $a$ as a function of $B$ includes the shift $\delta B_0$, $a=a_{\rm fs}(B-\delta B_0)$, as
already defined in (\ref{eq:acfb}).
This may introduce a systematic discrepancy
between our theory curves and the experimental data in Fig.\ref{fig:F}, in the form of shifts
of the abscissas of the experimental points. We thus define the horizontal shift
\bea
\delta\left(\frac{1}{k_F a}\right) &\equiv& \frac{1}{k_F a_{\rm fs}(B-\delta B_0)} - \frac{1}{k_F a_{\rm fs}(B)}
\\
&=& 
\frac{\delta B_0}{\Delta B} \, \frac{1}{k_F a_{\rm bg}}
\frac{(1-a_{\rm bg}/a)^2}{1-(1-a_{\rm bg}/a)\delta B_0/\Delta B}.
\label{eq:shift_gen}
\eea
To leading order
in $\eta$ (that is for $|\delta B_0/\Delta B|\ll 1$),
and using the fact that the data from~\cite{Hulet} reproduced in Fig.~\ref{fig:F} 
is sufficiently close to resonance to have
$|1-a_{\rm bg}/a|< 1.4$, 
this simplifies to:
\be
\delta\left(\frac{1}{k_F a}\right) \simeq
\frac{\delta B_0}{\Delta B} \, \frac{1}{k_F a_{\rm bg}}
(1-a_{\rm bg}/a)^2.
\label{eq:shift_simple}
\ee
Using the parameters of the Rice experiment~\cite{HuletPrivateData} and Eq.(\ref{eq:dB0surDB0}) this gives
\be
\left|\delta\left(\frac{ 1}{k_F a}\right) \right| < \frac{|\eta|}{1000}.
\label{eq:shift<<1}
\ee
Even without knowing $\eta$, we can thus resonably expect that
this shift is negligible on the scale of Fig.~\ref{fig:F}, i.e. that
\be
\left|\delta\left(\frac{ 1}{k_F a}\right) \right| \ll 1.
\label{eq:condition_shift<<1}
\ee
This is also consistent with the fact that no systematic horizontal shift
is apparent in Fig.\ref{fig:F} between the experimental data and the theory curves.

\subsection{Effect of the position dependence of $B_{\rm eff}(\rr)$}
\label{subsec:Beff(r)}

We now discuss the effect on the many-body properties
of the gas of the inhomogeneity of $B_{\rm eff}(\rr)$, that is of
a spatial dependence of the scattering length $a(\rr)$, see (\ref{eq:a_de_r}),
now taking into account the $\rr$ dependence of $U(\rr)$.
We rewrite (\ref{eq:Beff(r)}) as
\be
B_{\rm eff}(\rr) = B+\frac{2\eta U(\rr)}{\mu_b}  = B - \delta B_0  
+\frac{2\eta [U(\rr)-U(\mathbf{0})]}{\mu_b}.
\ee
When inserted in (\ref{eq:a_de_r}) this leads to the following position dependence
for the scattering length:
\be
\frac{1}{a(\rr)-a_{\rm bg}} = \frac{1}{a-a_{\rm bg}} 
-\eta\frac{2 m R_*}{\hbar^2} [U(\rr)-U(\mathbf{0})],
\label{eq:arex}
\ee
where we recall that $a$ is the scattering length in the trap center.
Right on resonance, $|a|=+\infty$, 
$R_*$ is one of the contributions to the interaction
effective range, see (\ref{eq:re_res}),
so that the zero-range limit implies $k_F |R_*| \ll 1$. 
Since the typical value of $U(\rr)-U(\mathbf{0})$ over the cloud size
is of the order of $\hbar^2 k_F^2/2m$, we see that the small parameter
$k_F R_*$ appears in (\ref{eq:arex}) so that we may
expect that the position dependence of $a(\rr)$ is negligible 
in the zero-range limit (for a fixed value of $\eta$).

Let us investigate this expectation more quantitatively and beyond the unitary limit, 
however still in a limiting case with simplifying assumptions. 
First, we assume that we are sufficiently close to resonance to have $|a(\rr)|\gg
|a_{\rm bg}|$ over the atomic cloud size. This implies that the excursion
of $B_{\rm eff}(\rr)$ over the cloud size is small as compared to $|\Delta B|$;
since the excursion of $U(\rr)$ over the laser {\sl trap} size is at most
$|U(\mathbf{0})|$, we note that this condition is automatically satisfied 
for $|\delta B_0|\ll |\Delta B|$.
Second, we shall assume that the gas may be treated in the zero temperature
local density approximation.
Third, we use the quadratic approximation (\ref{eq:Uharm}) for $U(\rr)$,
further supposing for simplicity that the harmonic trap is
isotropic, a case to which one can always reduce within the local density approximation
by a rescaling of the coordinates, see appendix \ref{app:LDA}.
Equation (\ref{eq:arex}) then reduces to
\be
\frac{1}{a(r)} \simeq \frac{1}{a} -\eta \frac{m R_*}{\hbar^2} m\omega^2 r^2,
\label{eq:invar}
\ee
and the gas density profile $n(\rr)$ is given by
\be
\mu-U(\vn) = \frac{1}{2} m\omega^2 r^2 + \mu^{\rm hom}[n(\rr),\frac{1}{a(\rr)}]
\label{eq:lda_easy}
\ee
where $\mu^{\rm hom}[n,1/a]$ is the chemical potential of a free space homogeneous
gas of density $n$ and scattering length $a$.

Let us derive simple conditions to have a weak effect of the scattering length
inhomogeneity on the density profile. Some of these conditions will be expressed
in terms of $k_F$ defined by (\ref{eq:ef}).

In the BCS regime, $a<0$ and $k_F |a|<1$,
we have $\mu-U(\vn) \simeq k_B T_F =\hbar^2 k_F^2/2m$.
Over the Thomas-Fermi radius $R$ of the cloud, such that
$\frac{1}{2} m\omega^2 R^2 \simeq k_B T_F$, we see that the relative change of
$1/a(r)$ is negligible if
\be
|1-a/a(R)| \simeq  (k_F |a|) (|\eta| k_F R_*)  \ll 1.
\label{eq:condbcs}
\ee
This is satisfied in the zero-range limit, for fixed $\eta$, according 
to the condition (\ref{eq:cond_re}) written for $k=k_F$.

In the unitary limit $k_F |a|\gg 1$, the previous reasoning fails, because
requiring that the relative change of $a(r)$ is small is too stringent.
Taking for simplicity the limiting case $|a|=+\infty$, 
it is sufficient to check that the gas remains locally unitary except
in a small region near the edge of the cloud.
First, to zeroth order in $\eta$, we calculate the local Fermi wavevector
$k_F^{\rm hom}(\rr)$  using the unitary equation of state 
$\mu^{\rm hom}[n,1/a=0]=\xi \hbar^2 (k_F^{\rm hom})^2/2m$,
where $k_F^{\rm hom}$ is defined in (\ref{eq:defkhom}), and using
the value of the chemical potential $\mu-U(\vn)=\sqrt{\xi} k_B T_F$.
We obtain
\be
k_F^{\rm hom}(r) = \frac{k_F}{\xi^{1/4}} \left(1-\frac{r^2}{R^2}\right)^{1/2}
\label{eq:kfhom_ex}
\ee
where $R$ is the unperturbed Thomas-Fermi radius such that $m\omega^2 R^2/2
=\sqrt{\xi} k_B T_F$.
Then, we include the $\eta$ position dependent term in $1/a(r)$
to calculate at which distance $l$ from the edge of the cloud 
the gas starts leaving the unitary regime, that is 
\be
k_F^{\rm hom}(R-l) |a(R-l)|=1.
\ee
Assuming {\sl a priori} $l\ll R$ we obtain from (\ref{eq:kfhom_ex})
\be
\frac{l}{R} \simeq 
\frac{\xi^{3/2}}{2} (\eta k_F R_*)^2.
\ee
We conclude {\sl a posteriori}
that $l/R \ll 1$ and that the spatial dependence of $a(r)$
is negligible when
\be
|\eta| k_F R_* \ll 1.
\label{eq:condunit}
\ee
As in the BCS case, this is equivalent for a fixed $|\eta|$ to the zero range
condition (\ref{eq:cond_re}) written here with $k=k_F$ and $|a|=+\infty$.

Finally we turn to the BEC regime $a>0$ and $k_F |a|<1$. 
We then use the approximate equation of state
\be
\mu^{\rm hom}[n(r),1/a(r)] \simeq -\frac{\hbar^2}{2 m a^2(r)} + 
\frac{\pi\hbar^2}{2m} \alpha a(r) n(r)
\ee
where $\alpha\simeq 0.60$ is the proportionality factor in (\ref{eq:ad})
between the dimer-dimer scattering length $a_d$ and $a$.
In particular, this requires $a\gg R_*$, see (\ref{eq:simple}).
Replacing $1/a(r)$ by (\ref{eq:invar})
in the above expression, we find that the leading term
in $1/a^2(r)$ gives rise to an external potential that sums up with the trapping potential
to give 
\bea
U_{\rm eff}(r) &\equiv& U(r) - \frac{\hbar^2}{2 m a^2(r)} \\
&=& U_{\rm eff}(0) +\frac{1}{2} \omega_{\rm eff}^2 r^2 
-\frac{m (\eta R_*)^2}{2\hbar^2}(m\omega^2 r^2)^2.
\eea
We see that the atomic oscillation frequency is renormalized into
\be
\omega_{\rm eff} = \left(1+\frac{2\eta R_*}{a}\right)^{1/2} \omega.
\ee
To have a small effect of the scattering length inhomogeneity 
on the trapped gas, we first require $\omega_{\rm eff}\simeq \omega$ so that
\be
|\eta| R_*/a \ll 1,
\label{eq:dure}
\ee
a condition with no counterpart in the BCS regime and more stringent than
(\ref{eq:condunit}) since $k_F a <1$ here. Again, we find that we recover, for
a fixed $|\eta|$, the zero range condition (\ref{eq:cond_re}) written here
for $k=1/a$.
Second we require that the $r^4$ term in $U_{\rm eff}(r)$ remains small as
compared to the $r^2$ one over the Thomas-Fermi radius $R$ of the condensate
of dimers, calculated here to zeroth order in $\eta$; we reach the
second condition
\be
\frac{m (\eta R_*)^2}{\hbar^2} m\omega^2 R^2 \simeq (\eta k_F R_*)^2 
\left(\frac{5}{64}k_F a_d\right)^{2/5} \ll 1.
\ee
Since $k_F a <1$ here, this second condition is automatically
ensured by the first condition (\ref{eq:dure}).
Third, we require that the relative variation of $1/a(r)$ is small
over the Thomas-Fermi radius $R$ of the condensate, to ensure that the 
spatially inhomogeneity of $a(r)$ in the mean field term
$[\pi\hbar^2/(2m)]\alpha a(r) n(r)$ has a small impact on the density
profile:
\be
|1-a/a(R)|\simeq (k_F a) (|\eta| k_F R_*) 
\left(\frac{5}{64}k_F a_d\right)^{2/5} \ll 1,
\ee
which is also implied by (\ref{eq:dure}).

To summarize, discussing perturbatively the effect of the position dependence
of $B_{\rm eff}(\rr)$ on the gas density profile in the local density
approximation, we find close to the Feshbach resonance ($|a|\gg |a_{\rm bg}|$, and also
$a\gg R_*$ in the BEC regime)
that the effect is small under 
the condition (\ref{eq:condbcs}) in the BCS regime,
under the condition (\ref{eq:condunit}) in the unitary regime,
and under the condition (\ref{eq:dure}) in the BEC regime: In all three cases,
this is the zero range condition (\ref{eq:cond_re}), written respectively for
$k=k_F, k=k_F$ and $k=1/a$,
and {\sl multiplied} by the parameter $|\eta|$.
If $|\eta|$ is not too large, neglecting the $r$ dependence of the scattering length
is thus automatically justified in the zero-range limit.

Interestingly,  in the unitary limit, it is possible to relate $\eta k_F R_*$ to the quantity
$\delta(1/k_Fa)$ introduced in the previous subsection:
\be
|\eta| k_F R_* \xi^{1/2} \simeq 
\left|\delta\left(\frac{1}{k_F a}\right) \frac{\mu-U(\vn)}{U(\vn)}\right|.
\label{eq:couche}
\ee
Experimentally, the gas is confined at the bottom of the laser trap 
so that $[\mu-U(\vn)]/|U(\vn)|\ll1$ 
(its value near unitarity at Rice is $0.15$~\cite{HuletPrivateData}). 
Together with the assumption (\ref{eq:condition_shift<<1}) 
this implies that the right-hand-side of (\ref{eq:couche}) is
$\lll 1$, and that the position dependence of $a(\rr)$ has a negligible effect
on the gas density.

A quantitative extension of the present qualitative discussion is to evaluate
the density change $\delta n(r)$ of the gas to first order in $\eta$.
To this end, one simply expands (\ref{eq:lda_easy}) to first order in $\eta$ around
the solution $n_0(r)$ with $\eta=0$, writing to first order
$n(r)=n_0(r)+\delta n(r)$ and $\mu=\mu_0+\delta\mu$:
\be
\delta n(\rr) = \frac{\delta\mu - \delta\left(\frac{1}{a}\right)(r)
\partial_{1/a}\mu^{\rm hom}[n_0(r),1/a]}{\partial_n \mu^{\rm hom}[n_0(r),1/a]}
\ee
where we have introduced the variation of $1/a(r)$ around $\eta=0$
to first order in $\eta$, not supposing $|a(r)|\gg |a_{\rm bg}|$ in
(\ref{eq:arex}):
\be
\delta\left(\frac{1}{a}\right)(r) = - \left(1-\frac{a_{\rm bg}}{a}\right)^2
\eta \frac{m R_*}{\hbar^2} m\omega^2 r^2.
\ee
$\delta \mu$ is then determined by the condition of a fixed number of particles
$\int_{r<R_0} d^3r \, \delta n(r)=0$,
where $R_0$ is the Thomas-Fermi radius of the gas for $\eta=0$.
In this way,
taking the variation of (\ref{eq:dlg}) to first order in $\eta$ around $\eta=0$,
one can calculate the effect of a non-uniformity of $a(r)$
on $dE/d(1/a)$ to first order in $\eta$.
We give here the result for the unitary gas:
\begin{multline}
\frac{\delta\left[dE/d(1/a)\right]}{dE/d(1/a)|_{\eta=0}} =
\eta k_F  R_* \left[
\left(\frac{64}{25\pi}-\frac{63\pi}{256}\right) \xi^{-1/4}\zeta \right. \\
\left.
+\frac{315\pi}{2048}\, \xi^{3/4}\zeta^{-1} f^{(2)}(0)
+\frac{2}{3}\, \xi^{1/2} k_F a_{\rm bg}\right],
\label{eq:resunit}
\end{multline}
where we recall that $\xi=f(0)$, $\zeta=-f'(0)$ , see (\ref{eq:expan}),
and $f^{(2)}(0)$ is the second order derivative of $f(x)$ with respect
to $x$ in $x=0$.
Note that, in this first order variation with respect to $\eta$,
the trapping potential $U(\rr)$ is kept fixed so that $k_F$ is also fixed.

The calculation (\ref{eq:resunit}) shows that, at unitarity, 
a measurement of the variation
of the number of closed-channel molecules $N_b$ as a function of 
$U_b(\rr)/U(\rr)$ may give access to $f^{(2)}(0)$.
It also shows that, in the unitary limit,
the condition to neglect $a_{\rm bg}$ with respect to $a(\rr)$ 
as done in the qualitative reasoning around (\ref{eq:condunit}),
is $k_F |a_{\rm bg}| \ll 1$.

\subsection{Metastability issues for a position dependent scattering length}

We now discuss whether it is energetically possible for a pair of atoms to
form a dimer which could
 escape from the trap by tunneling.
The energy variation in such a process is
\be
\Delta E = -\frac{\hbar^2}{m a^2(\infty)} - 2\mu
\ee
where $\mu$ is the chemical potential of the trapped gas and $a(\infty)$ is the effective scattering length at infinity, which we  assume to be positive.
We discuss under which conditions this process is energetically forbidden, i.e. $\Delta E>0$.

We assume that the gas has the usual Thomas-Fermi density profile and is weakly affected by the position dependence of $B_{\rm eff}(\rr)$, see subection~\ref{subsec:Beff(r)}.
We distinguish between two regimes for the scattering length $a$ in the trap center: the negative-$a$ regime where $1/a\leq0$,
and the BEC regime where $0<k_F a<1$.
Under the usual assumption that the gas is confined in a small region around the bottom of the laser trap we have $\mu\simeq U(\vn)$ in the negative-$a$ regime and $\mu\simeq U(\vn)-\hbar^2/(2m a^2)$ in the BEC regime.
We also make the simplifying assumption $a(\infty)\gg|a_{\rm bg}|$ so that (\ref{eq:arex}) reduces to
\be
\frac{1}{a(\infty)}\simeq\frac{1}{a}+2\eta\frac{m}{\hbar^2}R_* U(\vn).
\label{eq:a(infini)}
\ee
In the BEC regime one easily deduces
\be
\frac{\Delta E}{2|U(\vn)|}\simeq 1 + \mathcal{A} - \mathcal{B}
\ee
where
\bea
 \mathcal{A}&\equiv&2\eta\frac{R_*}{a}
\\
 \mathcal{B}&\equiv& 2\eta^2 \frac{m R_*^2}{\hbar^2}|U(\vn)|;
\eea
since we already assumed (\ref{eq:dure}) we automatically have $|\mathcal{A}|\ll1$.
In the negative-$a$ regime, Eq.(\ref{eq:a(infini)}) immediately gives an upper bound on $1/a(\infty)$, from which we get
\be
\frac{\Delta E}{2|U(\vn)|} \gtrsim 1- \mathcal{B}.
\ee
In both regimes we conclude that under the condition
\be
|\mathcal{B}|\ll1
\ee
we have $\Delta E\gtrsim2|U(\vn)|$ so that the process is energetically forbidden.

Finally we note that under the assumption $|a|\gg|a_{\rm bg}|$ 
we have from (\ref{eq:shift_gen},\ref{eq:dB0surDB0})
\be
\delta\left(\frac{1}{k_F a}\right) \simeq \eta k_F R_* \frac{|U(\vn)|}{k_B T_F}
\ee
so that
\be
\mathcal{B}\simeq\left|\delta\left(\frac{1}{k_F a}\right)\right|^2 \frac{k_B T_F}{|U(\vn)|}.
\ee
In a typical experiment such as the one at Rice we have $k_B T_F<|U(\vn)|$; thus the condition $|\mathcal{B}|\ll1$ is satisfied as soon as $|\delta(1/k_F a)|\ll1$, which is likely to hold at Rice, see (\ref{eq:shift<<1}).

\section{Calculation of $dE/d(1/a)$ in the local density approximation}
\label{app:LDA}

Restricting to a spin balanced gas, 
we first consider the general case of the two-channel model (\ref{eq:definition_H2}), 
in which $U_b(\rr)\neq 2 U(\rr)$. As explained in the appendix \ref{app:Beff},
where an effective position dependent magnetic field $B_{\rm eff}(\rr)$ is introduced,
this leads to a position dependent scattering length $a(\rr)$.
As a consequence, in the local density approximation, here at zero temperature,
the gas density $n(\rr)$ is given by
\be
\mu = U(\rr) + \mu^{\rm hom}[n(\rr),1/a(\rr)]
\label{eq:lda_gen}
\ee
where $\mu^{\rm hom}[n,1/a]$ is the chemical potential of the free space homogeneous
gas of density $n$ and scattering length $a$.
The mean energy and particle numbers in the gas are given in the same approximation by
\bea
E &=& \int d^3r \, \{U(\rr)+\epsilon^{\rm hom}[n(\rr),1/a(\rr)]\}\, n(\rr) \\
N &=& \int d^3r \, n(\rr)
\eea
where $\epsilon^{\rm hom}[n,1/a]$ is the mean energy per particle of the free space gas.
We wish to calculate the derivative of $E$ with respect to $1/a$, where
$a$ is the scattering length in the trap center $\rr=\vn$, with the constraint
that $N$ should be fixed. 

Let us perform an infinitesimal variation $\delta\left(\frac{1}{a}\right)$
of $1/a$.  This leads to a variation $\delta\mu$ of the chemical potential 
and $\delta n(\rr)$
of the density, that may be calculated from (\ref{eq:lda_gen}) but this is not required here.
This also leads to a variation of $1/a(\rr)$, according to (\ref{eq:arex}):
\be
\delta\left(\frac{1}{a(\rr)}\right) = \left(\frac{1-a_{\rm bg}/a(\rr)}{1-a_{\rm bg}/a}\right)^2 
\delta\left(\frac{1}{a}\right).
\ee
The variation of the gas energy is then
\begin{multline}
\delta E = \int d^3r \,\, \delta n(\rr) \left[U(\rr)+\epsilon^{\rm hom}[n(\rr),1/a(\rr)]
\phantom{\frac{1}{1}}\right. \\
\left. +n(\rr)\frac{\partial \epsilon^{\rm hom}}{\partial n}[n(\rr),1/a(\rr)]\right] \\
 + \delta\left(\frac{1}{a(\rr)}\right)
n(\rr)\frac{\partial \epsilon^{\rm hom}}{\partial(1/a)} [n(\rr),1/a(\rr)].
\label{eq:interm}
\end{multline}
Writing $\epsilon^{\rm hom}[n,1/a]=E^{\rm hom}[N,V,1/a]/N$, where
$V$ is the volume and $N$ the particle number of the homogeneous gas, we find the relation
\be
n\, \frac{\partial \epsilon^{\rm hom}}{\partial n}[n,1/a]  = \mu^{\rm hom}[n,1/a]
-\epsilon^{\rm hom}[n,1/a].
\ee
Using this expression and (\ref{eq:lda_gen}) one finds that a simplification occurs in
(\ref{eq:interm}). Furthermore, the fact that the particle number is fixed imposes
\be
\int d^3r\, \delta n(\rr)=0
\ee
so that (\ref{eq:interm}) leads to
\begin{multline}
\left(\frac{dE}{d(1/a)}\right)_N =  \int d^3r\, n(\rr) 
\left(\frac{1-a_{\rm bg}/a(\rr)}{1-a_{\rm bg}/a}\right)^2 \\
\times \frac{\partial \epsilon^{\rm hom}}{\partial(1/a)} [n(\rr),1/a(\rr)].
\label{eq:dlg}
\end{multline}
We note that an alternative derivation of this result is possible: One can introduce the density
of closed-channel molecules $n_b^{\rm hom}[n,1/a]$ for a free space gas of density $n$
and scattering length $a$, a quantity that can be related to the derivative of $\epsilon^{\rm hom}[n,1/a]$
with respect to $1/a$ simply by applying the general formula (\ref{eq:nb}) to a free space
system. Then setting for the trapped system in the local density approximation
$N_b=\int d^3r\, n_b^{\rm hom}[n(\rr),1/a(\rr)]$
and using (\ref{eq:nb}) for the trapped system leads to (\ref{eq:dlg}).

In practice, in the present paper, we perform explicit calculations
in the limiting cases presented in \S\ref{subsec:ariatilc}:
One takes the zero-range limit, which also allows to neglect
the spatial variation of $a(\rr)$ under conditions discussed in
the appendix \ref{app:Beff}.
Then $\epsilon^{\rm hom}$ is given by the relation (\ref{eq:def_f})
involving some universal function $f(x)$.
For the free space chemical potential we then introduce the convenient parametrization
\be
\mu^{\rm hom}[n,1/a]= \frac{\hbar^2}{2ma^2}
u\left(\frac{1}{k_F^{\rm hom}a}\right)
\ee
where $k_F^{\rm hom}=(3\pi^2n)^{1/3}$ is the ideal gas Fermi wavector,
and the function $u$ is
\be
u(x) = \frac{f(x)-\frac{1}{5}x f'(x)}{x^2}.
\ee
One also restricts to a harmonic trap for $U(\rr)$, see (\ref{eq:Uharm}).
The surface $n(\rr)=0$ is then an ellipsoid with Thomas-Fermi radii $R_\alpha$
along the directions $\alpha$. 

For $a<0$ we set
\be
\mu = \frac{\hbar^2 q^2}{2m}
\ee
with $q>0$. Then $\omega_\alpha R_\alpha = \hbar q/m$, and the density depends
only on  $X\geq 0$ such that $X^2=\sum_\alpha r_\alpha^2/R_\alpha^2$.
Equation (\ref{eq:lda_gen}) then reduces to the implicit equation
\be
q^2 a^2 (1-X^2) = u\left[\frac{1}{k_F^{\rm hom}(X) a}\right]
\ee
that one can solve numerically, using the interpolation for $f(x)$ given in appendix
\ref{app:spline}. 

For $a>0$, $\mu^{\rm hom}[n,1/a]$ tends to $-\hbar^2/(2m a^2)$ for $n\to 0$ so that we set
\be
\mu = \frac{\hbar^2 q^2}{2m} -\frac{\hbar^2}{2 m a^2}
\ee
with $q>0$. Then again $\omega_\alpha R_\alpha = \hbar q/m$, and (\ref{eq:lda_gen})
reduces to
\be
q^2 a^2 (1-X^2) = 1 + u\left[\frac{1}{k_F^{\rm hom}(X) a}\right]
\ee
to be solved numerically in general.

One can then calculate $dE/d(1/a)$ and $k_F$ defined in
(\ref{eq:ef}) from the expressions valid on both sides of the resonance:
\bea
\frac{dE}{d(1/a)} &=& \frac{3N\hbar^2}{10 m} \frac{\int_0^1 dX\, X^2 k_F^{\rm hom}(X)^4
f'\left[\frac{1}{k_F^{\rm hom}(X)a}\right]}{\int_0^1 dX\, X^2 k_F^{\rm hom}(X)^3} \\
k_F &=& \frac{\sqrt{2q}}{\pi^{1/3}} \left[4\pi\int_0^1 dX\, X^2 k_F^{\rm hom}(X)^3\right]^{1/6}
\eea
The analytical calculations performed in the weakly interacting regimes
are also obtained from the above formulas.


\section{Interpolation formula for the homogeneous gas energy}
\label{app:spline}

We explain how an interpolation formula was constructed
for the function $f$ defined in (\ref{eq:def_f}),
which was then used to produce the solid line in Fig.\ref{fig:F}a
and Fig.\ref{fig:F}b.

On the BEC side of the resonance, we applied a cubic spline 
interpolation to the points in Table I of \cite{Giorgini} with values of
$x=1/(k_F^{\rm hom}a)$ equal to $0$, $1$, $2$, $4$ and $6$,
using the last column of Table I (the one said to give $E/N-\epsilon_b/2$)
and adding to the corresponding data the quantity $-\hbar^2/(2m a^2)$.
We added a home made point at $x=8$ where the value of $f(x)$ and its first
order derivative $f'(x)$
are obtained from the bosonic Lee-Huang-Yang expansion (\ref{eq:serie_hom_bec}).

On the BCS side of the resonance we used the empirical interpolation formula
of \cite{Salasnich}, 
\be
f_S(x)=\alpha_1-\alpha_2 \arctan[\alpha_3 x 
(\beta_1-x)/(\beta_2-x)]. 
\ee
Constraints on $\alpha_1,\alpha_2,\alpha_3$ and $\beta_1,\beta_2$
are obtained from the values $f(0)$, $f'(0)$ 
and from the large $|x|$ expansion $f(x)=1+c_1/x+c_2/x^2+O(1/x^3)$,
where the coefficients $c_1$ and $c_2$ are given in (\ref{eq:serie_hom_bcs}).
This differs from \cite{Salasnich} where the coefficient
$c_2$ was not used and $\beta_2$ was deduced from a fit
to the Monte~Carlo data of \cite{Giorgini} on the BCS side.
Another difference with \cite{Salasnich} 
is that we took the value $f'(0)=-0.95$, that we
extracted from \cite{Lobo} using the relation (\ref{eq:tan1}).
Also, we set the value of $f(0)$ to $0.41$, which is within the error
bars of \cite{Giorgini}, in order to reduce the spurious discontinuity
of the second order derivative of $f(x)$ in $x=0$
between the interpolation formula for $x<0$
and the spline for $x>0$.
All this leads to $\alpha_1\simeq 0.41,\alpha_2\simeq 0.3756,
\alpha_3\simeq 1.062, \beta_1\simeq 0.9043,\beta_2\simeq 0.3797$. 

A test of this interpolation formulation 
is to compare its prediction for $f'(-1)$
with the value $\simeq -0.13$ that one can extract from
\cite{Lobo} using (\ref{eq:tan1}); one gets the satisfactory
result $f_S'(-1)\simeq -0.14$.

%
%

\bibliography{felix}

\end{document}